# GALEX UV Color Relations for Nearby Early-type Galaxies

José Donas[1], Jean-Michel Deharveng[1], R. Michael Rich[2], Sukyoung K. Yi[3], Young-Wook Lee[3], Alessandro Boselli[1], Armando Gil de Paz[4], Samuel Boissier[1], Stéphane Charlot[5], Samir Salim[2], Luciana Bianchi[6], Tom A. Barlow[7], Karl Forster[7], Peter G. Friedman[7], Timothy M. Heckman[8], Barry F. Madore[9], D. Christopher Martin[7], Bruno Milliard[1], Patrick Morrissey[7], Susan G. Neff[10], David Schiminovich[11], Mark Seibert[7], Todd Small[7], Alex S. Szalay[8], Barry Y. Welsh[12], and Ted K. Wyder[7]

*Laboratoire d'Astrophysique de Marseille, Traverse du Siphon, Les Trois Lucs, BP 8, 13376 Marseille Cedex 12, France*

jose.donas@oamp.fr

## ABSTRACT

[1]Laboratoire d'Astrophysique de Marseille, Traverse du Siphon, Les Trois Lucs, BP 8, 13376 Marseille Cedex 12, France

[2]Department of Physics and Astronomy, University of California, Los Angeles, CA 90095

[3]Center for Space Astrophysics, Yonsei University, Seoul 120-749, Korea

[4]Departamento de Astrofísica, Universidad Complutense de Madrid, Madrid 28040, Spain.

[5]Institut d'Astrophysique de Paris, CNRS, 98 bis boulevard Arago, F-75014 Paris, France

[6]Center for Astrophysical Sciences, The Johns Hopkins University, 3400 N. Charles St., Baltimore, MD 21218

[7]California Institute of Technology, MC 405-47, 1200 East California Boulevard, Pasadena, CA 91125

[8]Department of Physics and Astronomy, The Johns Hopkins University, Homewood Campus, Baltimore, MD 21218

[9]Observatories of the Carnegie Institution of Washington, 813 Santa Barbara St., Pasadena, CA 91101

[10]Laboratory for Astronomy and Solar Physics, NASA Goddard Space Flight Center, Greenbelt, MD 20771

[11]Department of Astronomy, Columbia University, New York, NY 10027

[12]Space Sciences Laboratory, University of California at Berkeley, 601 Campbell Hall, Berkeley, CA 94720

We use GALEX/optical photometry to construct color-color relationships for early-type galaxies sorted by morphological type. We have matched objects in the GALEX GR1 public release and the first IR1.1 internal release, with the RC3 early-type galaxies having a morphological type $-5.5 \leq T < -1.5$ with mean error in $T < 1.5$, and mean error on $(B-V)_T < 0.05$. After visual inspection of each match, we are left with 130 galaxies with a reliable GALEX pipeline photometry in the far-UV and near-UV bands. This sample is divided into Ellipticals ($-5.5 \leq T < -3.5$) and Lenticulars ($-3.5 \leq T < -1.5$). After correction for the Galactic extinction, the color-color diagrams FUV–NUV vs. $(B-V)_{Tc}$ are plotted for the two subsamples. We find a tight anti-correlation between the FUV–NUV and $(B-V)_{Tc}$ colors for Ellipticals, the UV color getting bluer when the $(B-V)_{Tc}$ get redder. This relationship very likely is an extension of the color-metallicity relationship into the GALEX NUV band. We suspect that the main source of the correlation is metal line blanketing in the NUV band. The FUV–NUV vs $B-V$ correlation has larger scatter for lenticular galaxies; we speculate this reflects the presence of low level star formation. If the latter objects (i.e. those that are blue both in FUV–NUV and $B-V$) are interpreted as harboring recent star formation activity, this would be the case for a few percent ($\sim 4\%$) of Ellipticals and $\sim 15\%$ of Lenticulars; this would make about 10 % of early–type galaxies with residual star formation in our full sample of 130 early-type galaxies. We also plot FUV–NUV vs. the $Mg_2$ index and central velocity dispersion. We find a tight anti-correlation between FUV–NUV and the $Mg_2$ index; we suspect this reflects blanketing in the NUV band being correlated with overall metallicity. We find a marginal anti-correlation of FUV$-V_T$ with $Mg_2$ for elliptical galaxies.

*Subject headings:* galaxies: elliptical and lenticular, cD — galaxies: photometry — galaxies: stellar content — ultraviolet: galaxies

## 1. Introduction

The far-ultraviolet radiation from early-type galaxies, first detected by the *Orbiting Astronomical Observatory-2* in 1969 (Code & Welch 1979), is now thought to be produced mainly by low-mass, helium-burning stars in extreme horizontal branch (EHB) and subsequent phases of evolution. O'Connell (1999) has extensively reviewed the data and the theoretical ideas that contributed to this interpretation.

This minority population of evolved stars has attracted attention beyond the domain of ultraviolet observations for several reasons. First, their characterization shed light on poorly-known phases of stellar evolution and associated astrophysical processes (giant-branch mass loss, helium enrichment) that are otherwise difficult to investigate. Second, their evolution has to be understood for more precise modeling of old stellar populations (Charlot et al. 1996) and for using the look-back time of their appearance as constraints on the age of elliptical galaxies (e.g. Greggio & Renzini 1990; Bressan et al. 1996; Yi et al. 1999; Lee et al. 2005). Last, because they are hot, these evolved stars may hide young stars that may trace departure from purely passive evolution, a long-sought feature in the discussion of monolithic vs. hierarchical scenarios of galaxy formation.

The most recent advances on the observational front have come from UV images with HST that, at the distance of M32 and the bulge of M31, resolved some of the UV emission into bright stars (Brown et al. 1998; 2000). Recently, a deep UV color-magnitude diagram with HST/STIS has confirmed a well-populated EHB in M32 but found fewer post-AGB stars than would have been expected in that population (Brown 2004), revealing a possible discrepancy with current evolution theory. Albeit extremely powerful, such direct observation of individual stars are limited to very few objects and will stay so for the foreseeable future.

A large sample of early-type galaxies is reachable only through integrated light measurements. Following the early observations reviewed by O'Connell (1999), such a possibility is now offered by the extensive survey made with the GALEX mission (Martin et al. 2005). This survey provides two UV broad band measurements, the far-UV at $\lambda_e = 1528$Å (FUV) and the near-UV at $\lambda_e = 2271$Å (NUV). The importance of having a near-UV band like the NUV band of GALEX has been addressed by Fanelli et al. (1988), Dorman et al. (2003) and its potential for revealing a young stellar component illustrated by Ferreras & Silk (2000) with HST observations of a cluster at $z = 0.41$. In a galaxy with no current star formation, the near-UV light near 2000Å is dominated largely by turnoff and subgiant stars, with some radiation from blue horizontal branch (HB) stars if present. The far-UV radiation at 1500Å and shortward is dominated by hot evolved stars, but the nature of these stars remains unsettled. The UV light is of great interest because the UV wavelengths are most sensitive to trace "residual" star formation in elliptical galaxies. In addition to being sensitive to residual star formation at the present epoch, the NUV band is particularly sensitive to the normal passive evolution of a stellar population with lookback time. Color-color diagrams are therefore powerful diagnostics of evolution, residual star formation, and UV-bright populations of evolved stars.

UV observations of elliptical galaxies with GALEX have already been reported. Lee et al. (2005) have addressed the issue of observing the UV emission in clusters at different redshifts. Rich et al. (2005) and Yi et al. (2005) have studied samples of early-type galaxies where the GALEX photometry has been matched with galaxies in the SDSS main galaxy sample. The first paper concentrates on the far-UV–optical color in quiescent early-type galaxies for studying the population responsible for the UV rising flux, the second on the near-UV color-magnitude relation, the scatter of which is interpreted in terms of residual star formation (see also Kaviraj et al. 2006; Schawinski et al. 2006). Boselli et al. (2005) have studied a sample of early-type galaxies with a large range of luminosity in the Virgo Cluster. Last, early-type galaxies are included in the GALEX Nearby Galaxy Survey (NGS) of Gil de Paz et al. (2006)

Following the completion of the GALEX survey over a significant fraction of the sky, this paper can now present the UV observations of nearby early-type galaxies (elliptical/lenticular), with a morphological type selection based on the Third Reference Catalog of Bright Galaxies (RC3) (de Vaucouleurs et al. 1991) and an emphasis on UV-optical color-color diagrams.

## 2. Data

### 2.1. GALEX observations

The Galaxy Evolution Explorer (GALEX) is a NASA Small Explorer mission launched by a Pegasus-XL vehicle on 2003 April 28 into a 690 km circular orbit. The instrument is designed to image a 1°.2 diameter circular field of view with its resolution of $\approx 4.5$" (FUV), 6.0" (NUV) set by detector electronics, and to perform imaging or spectroscopy in two ultraviolet bands simultaneously, far-UV (1344–1786Å) and near-UV (1771–2831Å). The imaging surveys include an all-sky imaging survey (AIS, $m_{AB} \simeq 20.5$), a medium imaging survey of 1000 deg$^2$ (MIS, $m_{AB} \simeq 23.0$), a deep imaging survey of 100 deg$^2$ (DIS, $m_{AB} \simeq 25.0$), and a nearby galaxy survey (NGS). The GALEX instrument and mission are described by Martin et al. (2005) and Morrissey et al. (2005). All observations are performed in a pointed mode with an arc-minute spiral dither. From the time-tagged photon lists, the data analysis pipeline generates integrated images. The data pipeline includes a photometric module which uses the SExtractor program (Bertin et al. 1996) for detection and photometry of the sources in the imaging data, with some modifications in the determination of sky background. The GALEX pipeline makes a few different measurements of the UV flux of a source. In the following we use the UV flux measured through 7 circular apertures with fixed diameters (3″.0, 4″.5, 7″.5, 12″.0, 18″.0, 25″.5 and 34″.5), and the UV flux measured through two elliptical apertures whose major axis $D_{NUV}$ and $D_{FUV}$ are scaled to 2.5 times the Kron diameter (Kron 1980) determined in the near-UV and far-UV images respectively.

GALEX uses the AB magnitude system of Oke & Gunn (1983), with the FUV and NUV magnitude defined by:

$$m_{\rm UV} = m_0 - 2.5 \log f_{\rm UV}$$

where $f_{\rm UV}$ is the count rate (flat-field corrected), and $m_0$ is the zero point ($m_0 = 20.082$ and 18.817 for NUV and FUV respectively) corresponding to the AB magnitude of 1 count s$^{-1}$ flat-field-corrected detection. The uncertainty on the GALEX zero points is estimated to be $\pm 0.15$ mag for both the far-UV and near-UV channels.

## 2.2. Morphological selection in the RC3 catalog

The RC3 catalog is reported to be reasonably complete for galaxies having apparent diameters larger than 1 arc-min at the $D_{25}$ isophotal level[1] and total B-band magnitudes $B_T$ brighter than about 15.5, with a redshift not in excess of 15,000 km s$^{-1}$. Additional objects meeting only the diameter or the magnitude condition, and objects of interest smaller than 1 arc-min, fainter than 15.5, or with redshifts $> 15,000$ km s$^{-1}$, are also included in the RC3. The total of RC3 objects is 23,022, among them 1103 have $(B-V)_T$ measurement[2], and are classified as Ellipticals or Lenticulars/S0, which correspond to values of the numerical index $-5.5 \leq T < -0.5$.

In order to obtain a reliable subsample of early-type galaxies, we select only the sources with morphological type in the range $-5.5 \leq T < -1.5$ with mean error on $T < 1.5$, and mean error on $(B-V)_T < 0.05$; this leads to an RC3 subsample of 875 galaxies.

## 2.3. Resulting sample

Among the RC3 subsample of 875 early-type galaxies, 138 (i.e $\sim 16\%$) are in the fields observed in the far-UV and near-UV bands by GALEX in the AIS, MIS, and NGS surveys included in the GR1 public release and the first IR1.1 internal releases.

Because the coordinates in the RC3 are not very accurate, we have searched for UV counterparts within a radius of $1.'5$, and examined individually each field in the far-UV, near-UV, and POSS-I E images; this is possible because the objects are relatively bright and in small number. After visual inspection we find 130 galaxies with a reliable GALEX pipeline photometry; 6 have doubtful photometry because of serious blend with close sources (PGC42620, 69260, 26377, 2569, 13457, 14375), and 2 are not detected in far-UV because of short exposures (PGC65580, 68241). Since 130 of the 138 galaxies have reliable UV photometry, we conclude that our sample is not UV magnitude-limited and could indeed be representative of the RC3 magnitude- and diameter-limited sample. The 130 early-type galaxies of our resulting sample are listed in Table 1 with some relevant information.

---

[1] $D_{25}$ is the apparent major isophotal diameter measured at or reduced to surface brightness level $\mu_B = 25.0$

[2] $(B-V)_T$ is the total (asymptotic) color index in the Johnson B–V system

Our sample includes $\sim 15\%$ of the RC3 selected early-types, without strong biases in the distribution of morphological types, $B_T$ magnitudes, $(B-V)_T$ color, or $D_{25}$ diameter (Fig.1). We conclude that our sample is relatively well representative of the early-type galaxies in the RC3.

The apparent magnitudes were corrected for galactic extinction using the Schlegel et al. (1998) reddening maps and assuming the extinction law of Cardelli et al. (1989). The resulting ratio of the extinction in the GALEX bands to the color excess $E(B-V)$ are $A_{FUV}/E(B-V) = 8.376$ and $A_{NUV}/E(B-V) = 8.741$ (Wyder et al. 2005). The mean color excess due to the Galactic extinction is $E(B-V) \simeq 0.04$. The optical color $(B-V)_T$ is converted in the AB system and corrected for Galactic extinction by $(B-V)_{Tc} = (B-V)_T - 0.119 - E(B-V)$. Since the median redshift of our sample is $z \sim 0.01$ no $K$-corrections are applied.

## 3. Analysis

### 3.1. FUV–NUV versus $(B-V)_{Tc}$ color-color diagram

We have chosen to display our data in a FUV–NUV vs. $(B-V)_{Tc}$ color-color diagram (Fig.2). This choice, instead of a color-magnitude diagram allows us to avoid significant dispersion due to uncertainties in distance (our sample of nearby galaxies span a redshift range $0 < z < 0.045$). The choice of the FUV–NUV color, based on GALEX measurements alone, minimize instrumental inhomogeneities. The FUV–NUV color is calculated using the FUV and NUV magnitudes (Table 1) measured by the GALEX pipeline through the elliptical aperture whose major axis is scaled to 2.5 times the Kron diameter *determined in the near-UV images*. We emphasize the UV color is calculated in the same aperture. The major diameter $D_{NUV}$ of this aperture (Table 1) is similar to the $D_{25}$ values given in the RC3.

Our sample of early-type is divided into 56 Ellipticals (defined as galaxies with morphological type $-5.5 \leq T < -3.5$) and 74 Lenticulars (defined as galaxies with morphological type $-3.5 \leq T < -1.5$). The color-color diagram clearly shows a tight anti-correlation between the FUV–NUV and $(B-V)_{Tc}$ colors for elliptical galaxies: the UV color decreasing when $(B-V)_{Tc}$ increases (Fig.2a). The amplitude range in the UV color is 2 magnitudes, against only 0.2 mag in $(B-V)_{Tc}$. When the two galaxies off the correlation are excluded the correlation coefficient is $r = -0.73$. This value reaches $-0.84$ if we select ellipticals with mean error on $(B-V)_{Tc} < 0.02$. For the early-type with morphological type in the range $-3.5 < T < -1.5$ which are lenticular galaxies (Fig.2b), the relation is more dispersed, and a larger fraction of objects are off the correlation.

Because elliptical galaxies exhibit a far-ultraviolet optical color gradient with the inner regions slightly bluer than the outer parts (e.g. Ohl et al. 1998; Rhee et al. 2006), we have examined how this gradient may affect the UV color in different apertures. The GALEX pipeline allows us to calculate the mean FUV–NUV color in 7 circular apertures with diameters fixed at 3.0, 4.5, 7.5, 12.0, 18.0, 25.5 and 34.5 arc-sec, and through two elliptical apertures whose major axis $D_{NUV}$ and $D_{FUV}$ are scaled to 2.5 times the Kron diameter determined in the near-UV and far-UV images respectively. As expected the mean FUV–NUV color becomes bluer when the aperture radius decreases. In our sample, the difference between the mean color in the $34''.5$ circular aperture and the mean color in the $D_{NUV}$ elliptical apertures is only 0.03 mag. There is no significant difference between the mean color when the UV fluxes are measured through the $D_{NUV}$ or $D_{FUV}$ elliptical apertures.

Gil de Paz et al. (2006) have presented the integrated photometry, surface-brightness and color profile for a total of 1034 nearby galaxies observed by GALEX. They obtain a main difference of $-0.19 \pm 0.20$ and $-0.23 \pm 0.20$ between the asymptotic magnitudes and the $D_{25}$ aperture magnitudes respectively for the FUV and NUV, resulting in a difference of 0.04 in the FUV–NUV color. Their sample includes 48 ellipticals ($-5.5 \leq T < -3.5$, and mean error on $T < 1.5$) and 49 lenticulars ($-3.5 \leq T < -1.5$, and mean error on $T < 1.5$) with FUV and NUV asymptotic magnitudes, and $B$, $V$ measurements. Among them, 73 are in common with our sample. The comparison of our FUV and NUV with asymptotic magnitudes measured by Gil de Paz shows a mean difference of $-0.06 \pm 0.27$ and $-0.19 \pm 0.20$ in far-UV and near-UV respectively. Our UV magnitudes are very close to the integrated aperture magnitudes in $D_{25}$ ellipses. The mean differences are $0.03 \pm 0.18$ for $NUV_{D_{25}}$–NUV and $0.06 \pm 0.21$ for $FUV_{D_{25}}$–FUV. The similarity of the $D_{25}$ and $D_{NUV}$ diameters explains the small differences between the $NUV_{D_{25}}$, $FUV_{D_{25}}$ aperture magnitudes and those used in this paper.

We conclude that the FUV–NUV colors obtained with large apertures or asymptotic photometry, are very similar for our sample. The trends in the FUV–NUV diagrams vs. $(B-V)_{Tc}$ are not significantly affected, even if the strength of the relations is slightly

– 8 –

changed (the correlation coefficient is −0.61 for the elliptical galaxies in common with the Gil de Paz et al. sample and asymptotic magnitudes).

### 3.2. UV color versus $Mg_2$ and Velocity dispersion

Using IUE observations of early-type galaxies, Faber (1983) and Burstein et al. (1988) found that the nuclear $(1550 - V)$ colors become bluer as the nuclear $Mg_2$ spectral line index increases. Burstein et al. (1988) also found significant correlations between $1550 - V$ and the central velocity dispersion. More recently, using a large sample of SDSS early-type galaxies observed by GALEX, Rich et al. (2005) reported the lack of a significant correlation between the (FUV–$r$) color and the Lick $Mg_2$ index or the velocity dispersion, while Boselli et al. (2005) found a mild trend between the FUV–NUV and $Mg_2$ in a sample of early-type galaxies in the Virgo Cluster.

We have plotted in Fig.3 the FUV–NUV and FUV$-V_T$ colors versus the $Mg_2$ index and the central velocity dispersion $\sigma_V$ for our sample. The $Mg_2$ index comes from the compilation of Golev & Prugniel (1998), and the $\sigma_V$ comes from the LEDA database [3].

Fig.3 (*top*) shows clearly that the FUV–NUV color of ellipticals become bluer as the spectral line index $Mg_2$ increases. The same trend appears between the UV color and the central velocity dispersion $\sigma_V$. It should be noted that the $Mg_2$ and $\sigma_V$ values given by Golev & Prugniel and the LEDA database are standardized to an aperture of 0.595 $h^{-1}$kpc (equivalent to an angular diameter of 3″.4 at the distance of Coma cluster). We have checked that when the FUV–NUV colors are calculated in a similar aperture (by linear interpolation between the UV flux measured by the GALEX pipeline through 7 circular apertures), the trends in Fig.3 (*top*) are not significantly affected. These correlations are opposite to the well-known dependence of $B - V$ color on metal abundance or central velocity dispersion, and are not surprising since, as showed in Section 3.1, there is an anti-correlation between the FUV–NUV and $B - V$ colors. If we consider that the velocity is a mass indicator, the more massive ellipticals tend to be bluer in FUV–NUV. If the velocity dispersion also correlates with the galaxy age (Thomas et al. 2005; Ziegler et al. 2005) the bluest in UV color would be the oldest. Lenticulars show a similar trend, but a larger fraction of galaxies are off the correlation.

The correlations found above with the FUV–NUV color (Fig.3 *top*) for the subsample of morphological ellipticals become marginal (not visible against velocity dispersion) with the FUV−$V_T$ color and suffer far greater scatter (Fig.3 *bottom*). For instance the correlation coefficient with the $Mg_2$ index decreases from −0.83 to −0.63 (NGC855 and NGC6482 excluded). This is expected because the FUV−$V_T$ color is not as homogeneous in terms of instrument and photometric aperture as the FUV–NUV color. The correlations with the FUV–$V_T$ color are not visible for the Lenticular subsample, suggesting that the selection of the sample is, in addition to the homogeneity of the UV color, a factor in the strength of the correlation.

---

[3]http://leda.univ-lyon1.fr

For the morphological ellipticals only, we recover an approximation of the correlation between FUV−$V_T$ and $Mg_2$ index that was seen by Burstein et al. (1988) in IUE data. However, Burstein also saw a correlation with velocity dispersion that we do not recover here, and the lenticular subset also fails to show a correlation. In comparison with Rich et al. 2005, our elliptical subset has been selected to be uniform in morphology, exhibits a tight color-color correlation, and covers a wide range in the $Mg_2$ index. These factors may have favored our detection of the mild correlation we report here. However, Rich et al. (2006), examining a large sample of nearby early-type galaxies, do not recover the Burstein et al. correlation; we suspect that strict morphological selection is required to recover even the mild version that we see here.

### 3.3. Discussion

Before any interpretation of the color-color diagrams of Fig.2, we have evaluated the possible UV flux contamination from AGN activity by measuring the contribution of the central part of the galaxies to the UV emission. This evaluation is possible because the major part of the galaxies in our sample are relatively well spatially resolved by GALEX. We have estimated the UV flux in the central $6''$ aperture (the typical angular resolution of GALEX), by linear interpolation between the UV flux measured by the GALEX pipeline through the $4''\!.5$ and $7''\!.5$ apertures. Among the 119 galaxies with $\log D_{25} > 1$ (i.e. major apparent diameter $> 60''$), we find 3 galaxies in the near–UV band and 8 in the far–UV band for which the central ($< 6''$) UV flux (stellar population + AGN) is $> 30\%$ of the total UV flux. These galaxies, likely candidates for AGN, are indicated in Fig.2 by open circles. The trends in the color-color diagrams are not affected by the possible AGN activity.

From Fig.2, it appears that morphological elliptical galaxies follow a much tighter anti-correlation between FUV–NUV and $B - V$ than the lenticular galaxies. We propose that we are seeing a metallicity-driven correlation, with the greater dispersion for the lenticulars arising from residual star formation. We suspect that blanketing in the NUV band (consider this band as an extended U-band) is the primary driver of this observed correlation. The light in the NUV bandpass is dominated by the hottest stars in an old stellar population, which would be turnoff and subgiant stars, or blue horizontal branch stars if present. Any young stellar population would make the FUV–NUV and $B-V$ colors simultaneously bluer, according to their current spectral energy distribution. It is also conceivable that factors such as age, metallicity, mass or a combination of these, currently advocated for the spread in the $B-V$ colors of early–type galaxies may redden the $B-V$ color while simultaneously decreasing the NUV and/or increasing the FUV radiation and making the FUV–NUV bluer. Fig.4 shows clearly a positive correlation between the NUV$-V_T$ and $B-V$ colors, and a marginal anti-correlation between the FUV$-V_T$ and $B-V$ colors for ellipticals. This is consistent with our suggestion that the stars contributing to the color-color correlation are likely old metal rich stars, with the dominant contributors in the NUV band being the turnoff and subgiant stars (e.g. Dorman, O'Connell & Rood 2003). The tight correlation for the ellipticals appears consistent with little intrinsic variation, from galaxy to galaxy, of the hot stellar populations (e.g. post-AGB or EHB stars) dominating the FUV band. The dominant driver in this color-color relationship appears to be metallicity and (in the case of lenticulars) star formation.

It is of interest to check how these results compare to the predictions of the stochastic library of model SF histories used by Salim et al. (2005) to interpret the 7-band photometry of combined sample of GALEX and SDSS galaxies. We note that this library is based on the Bruzual & Charlot (2003) population synthesis models, which were not optimized to describe the UV upturn of early-type galaxies (in these models, the UV light of old stellar populations is produced primarily by post-AGB rather than EHB stars). First, the anti-correlation between the FUV–NUV and $B-V$ colors is very well reproduced (Fig.5a) if we select models representative of old galaxies (mean age of star formation episodes in the 8–12 Gyr interval) with low extinction ($\tau_V < 0.5$). These galaxies formed the bulk of their stars at $z > 1$ (assuming $\Omega_M = 0.3, \Omega_\Lambda = 0.7, H_0 = 70$ kms$^{-1}$Mpc$^{-1}$). The spread in UV and $B-V$ colors can be easily reproduced by the variation of metallicity, galaxies with high metallicity values ($1.0 < Z/Z_\odot < 2$) being bluer in UV color (FUV–NUV$\lesssim 1.2$), while the galaxies with low metallicity ($0.2 < Z/Z_\odot < 1.0$) are redder (FUV–NUV$\gtrsim 1.2$). The models have old stellar populations but the only hot star contributors are post-AGB stars which do not have a metallicity dependence in the models and are not the contributors favoured by the most recent observations (Brown et al. 2000, Brown 2004). This suggests again that blanketing in NUV is presumably the main cause of the agreement between the models and the UV color-color correlation for ellipticals. The FUV flux would therefore be dominated by hot stars (whose Post-AGB in the models are only proxy) that would have low blanketing as suggested by the spectra of Brown et al. (1997). The stars providing most of the NUV flux would be those near the main sequence turnoff.

– 11 –

Secondly, the departure from the trend (or the simultaneous decrease of the FUV–NUV and $B-V$ colors) is well reproduced with the models with mean age of star formation episodes in the 4–12 Gyr interval (Fig.5b), and can be interpreted as recent star formation activity. We expect low impact of the internal extinction in the scatter in Fig.5 because the internal extinction is generally very small in most nearby early-type (Dorman, O'Connell & Rood 2003) galaxies and the FUV–NUV color is essentially reddening free for moderate amounts of reddening. In fact, the color-color correlation may prove to be a powerful detector of weak star formation activity. Fig.2 shows it may be the case for 4 % of the ellipticals and about 15 % of the lenticulars. This would make about 10 % of early–type galaxies with residual star formation in our full sample of 130 early-type galaxies. This number is lower than the fraction of 15 % reported by Yi et al. (2005) with a different approach based on a NUV color–magnitude relation and the sample of galaxies identified as early-type by Bernardi et al. (2003) in the SDSS. That the derivation of this number is depending on the selection criteria used to define the sample of early–type galaxies is shown directly by our approach; the fraction of early–type galaxies with sign of star formation increases from 4 % to 15 % when selecting the RC3 morphological types $-3.5 < T < -1.5$ instead of $-5.5 < T < -3.5$. It is likely that a number of the features leading to a non classification in the types $-5.5 < T < -3.5$ are somewhat related to star formation activity. At large distances, selection of early-type objects, based on concentration index, luminosity profile and spectra may naturally lead to a larger fraction of residual star formation. At the opposite, it should be possible to combine a color selection with the morphological selection of nearby galaxies that would retain only passively evolving early–type galaxies and no residual star formation.

Another illustration on how the fraction of early-type galaxies with residual star formation is depending on the selection of the objects is shown by Boselli et al. (2005); their sample of early-type galaxies in the Virgo cluster contains a large fraction of dwarf galaxies whose colors, different from those of giant ellipticals, are interpreted in terms of recent star formation activity.

## 4. Conclusions

From the cross-match of the GALEX observations with the RC3 catalog, we have obtained a reliable sample of 130 early-type galaxies, divided in two subsamples: 56 Ellipticals ($-5.5 \leq T < -3.5$) and 74 Lenticulars ($-3.5 \leq T < -1.5$). The main results are:

1. There is an anti-correlation between the FUV–NUV and $(B-V)_{Tc}$ colors (the UV color decreasing of two magnitudes when $(B-V)_{Tc}$ increases by 0.2 mag.) in the Elliptical subsample. This relation is more dispersed and a larger fraction of objects are off the correlation in the Lenticular subsample. We conclude that this correlation reflects an extension into the UV of the color-metallicity relationship for ellipticals and is mainly driven by metal line blanketing in the near-UV band. Most of the flux in this band would originate from stars near the main sequence turnoff while the FUV flux is dominated by hot evolved stars whose precise nature has yet to be determined.

2. There is a clear correlation between the FUV–NUV color and the $Mg_2$ spectral index (also velocity dispersion) for the Elliptical subsample. However, this correlation gets worse either with the FUV$-V_T$ color, or the Lenticular subsample (to the point of disappearing with a combination of both). We conclude that this correlation is an additional aspect of the color-metallicity relationship we have noted earlier.

3. The anti-correlation between FUV–NUV color and $B - V$ is especially tight for morphological ellipticals while showing more scatter for the lenticulars. We find this to be a sensitive means of detecting residual star formation; 10% of our sample shows evidence for residual star formation. We are able to model this effect using the Bruzual & Charlot (2003) models.

4. The trends found either diminish in quality (less tight relations, large number of outliers) or disappear from the Elliptical to the Lenticular subsample. Relevant properties of early-type galaxies are therefore crucially depending on how strictly is defined the sample of these objects.

The sample of 130 galaxies presented in this paper represents $\sim$ 15% of the reliable early-type of the RC3 catalog; when the GALEX survey is completed we can expect to multiply by a factor $\sim$ 6 the number of early-type galaxies in the sample.

We wish to acknowledge Jakob Walcher for kindly extracting for us useful data from stochastic realizations of SF models. GALEX (Galaxy Evolution Explorer) is a NASA Small Explorer, launched in April 2003. We gratefully acknowledge NASA's support for construction, operation, and science analysis for the GALEX mission, developed in cooperation with the Centre National d'Etudes Spatiales of France and the Korean Ministry of Science and Technology. The grating, window, and aspheric corrector were supplied by France. We acknowledge the dedicated team of engineers, technicians, and administrative staff from JPL/Caltech, Orbital Sciences Corporation, University of California, Berkeley, Laboratory Astrophysique Marseille, and the other institutions who made this mission possible. This work was partly supported by grant No. R01-2006-000-10716-0 (SKY) from the Basic Research Program of the Korea Science & Engineering Foundation. Gil de Paz, A., is financed by the MAGPOP EU Marie Curie Research Training Network.

*Facility:* GALEX

Table 1. The early-type galaxies sample

| PGC (1) | AName (2) | T (3) | $E(B-V)$ (4) | $(B-V)_{Tc}$ (5) | $logD_{25}$ (6) | $logD_{NUV}$ (7) | NUV (8) | FUV (9) | FUV$-V_T$ (10) | Mg$_2$ (11) | $log\sigma_V$ (12) |
|---|---|---|---|---|---|---|---|---|---|---|---|
| PGC34780 | NGC3641 | $-5.3 \pm 0.4$ | 0.042 | $0.74 \pm 0.02$ | 1.03 | 1.41 | $17.59 \pm 0.009$ | $19.13 \pm 0.042$ | $6.10 \pm 0.21$ | $0.28 \pm 0.006$ | $2.21 \pm 0.010$ |
| PGC02149 | NGC163 | $-5.0 \pm 0.4$ | 0.034 | $0.79 \pm 0.02$ | 1.19 | 1.22 | $18.15 \pm 0.013$ | $19.28 \pm 0.037$ | $6.73 \pm 0.15$ | – | $2.31 \pm 0.026$ |
| PGC03513 | UGC610 | $-5.0 \pm 0.8$ | 0.071 | $0.86 \pm 0.04$ | 1.26 | 1.08 | $19.18 \pm 0.151$ | $20.05 \pm 0.218$ | $7.36 \pm 0.30$ | – | – |
| PGC04376 | NGC430 | $-5.0 \pm 0.6$ | 0.030 | $0.83 \pm 0.03$ | 1.12 | 0.97 | $18.27 \pm 0.067$ | $19.54 \pm 0.157$ | $7.16 \pm 0.43$ | $0.28 \pm 0.009$ | $2.46 \pm 0.026$ |
| PGC05663 | NGC584 | $-5.0 \pm 0.3$ | 0.042 | $0.80 \pm 0.01$ | 1.62 | 1.58 | $15.81 \pm 0.004$ | $17.63 \pm 0.019$ | $7.33 \pm 0.13$ | $0.26 \pm 0.003$ | $2.32 \pm 0.009$ |
| PGC07584 | NGC777 | $-5.0 \pm 0.3$ | 0.047 | $0.87 \pm 0.02$ | 1.39 | 1.29 | $17.19 \pm 0.008$ | $17.70 \pm 0.020$ | $6.44 \pm 0.14$ | $0.33 \pm 0.008$ | $2.51 \pm 0.014$ |
| PGC08557 | NGC855 | $-5.0 \pm 0.7$ | 0.072 | $0.52 \pm 0.02$ | 1.42 | 1.38 | $15.31 \pm 0.004$ | $15.93 \pm 0.010$ | $3.61 \pm 0.13$ | $0.03 \pm 0.008$ | $1.85 \pm 0.058$ |
| PGC09890 | NGC990 | $-5.0 \pm 0.8$ | 0.112 | $0.80 \pm 0.02$ | 1.26 | 1.10 | $17.70 \pm 0.073$ | $18.90 \pm 0.139$ | $6.84 \pm 0.21$ | $0.24 \pm 0.005$ | $2.22 \pm 0.030$ |
| PGC10175 | NGC1052 | $-5.0 \pm 0.3$ | 0.027 | $0.79 \pm 0.01$ | 1.48 | 1.58 | $15.83 \pm 0.003$ | $16.85 \pm 0.010$ | $6.51 \pm 0.13$ | $0.30 \pm 0.005$ | $2.32 \pm 0.008$ |
| PGC13299 | NGC1379 | $-5.0 \pm 0.4$ | 0.012 | $0.76 \pm 0.01$ | 1.38 | 1.56 | $16.27 \pm 0.004$ | $18.05 \pm 0.017$ | $7.22 \pm 0.10$ | $0.25 \pm 0.003$ | $2.08 \pm 0.010$ |
| PGC13418 | NGC1399 | $-5.0 \pm 0.3$ | 0.013 | $0.83 \pm 0.01$ | 1.84 | 1.51 | $14.72 \pm 0.002$ | $15.20 \pm 0.004$ | $5.70 \pm 0.10$ | $0.33 \pm 0.003$ | $2.53 \pm 0.007$ |
| PGC13419 | NGC1395 | $-5.0 \pm 0.3$ | 0.023 | $0.82 \pm 0.01$ | 1.77 | 1.47 | $15.62 \pm 0.017$ | $16.44 \pm 0.037$ | $6.96 \pm 0.07$ | $0.31 \pm 0.003$ | $2.39 \pm 0.009$ |
| PGC13433 | NGC1404 | $-5.0 \pm 0.3$ | 0.011 | $0.84 \pm 0.01$ | 1.52 | 1.61 | $15.36 \pm 0.003$ | $16.41 \pm 0.009$ | $6.49 \pm 0.13$ | $0.31 \pm 0.003$ | $2.37 \pm 0.008$ |
| PGC13505 | NGC1407 | $-5.0 \pm 0.3$ | 0.069 | $0.84 \pm 0.01$ | 1.66 | 1.74 | $14.93 \pm 0.003$ | $15.64 \pm 0.010$ | $6.22 \pm 0.20$ | $0.32 \pm 0.004$ | $2.44 \pm 0.009$ |
| PGC13638 | NGC1426 | $-5.0 \pm 0.3$ | 0.017 | $0.76 \pm 0.01$ | 1.42 | 1.32 | $16.84 \pm 0.031$ | $18.30 \pm 0.083$ | $7.01 \pm 0.10$ | $0.24 \pm 0.006$ | $2.19 \pm 0.009$ |
| PGC13738 | NGC1439 | $-5.0 \pm 0.3$ | 0.030 | $0.73 \pm 0.01$ | 1.39 | 1.33 | $16.74 \pm 0.033$ | $18.44 \pm 0.094$ | $7.18 \pm 0.11$ | $0.27 \pm 0.009$ | – |
| PGC13814 | NGC1453 | $-5.0 \pm 0.4$ | 0.105 | $0.83 \pm 0.01$ | 1.38 | 1.15 | $17.03 \pm 0.062$ | $18.16 \pm 0.117$ | $7.00 \pm 0.18$ | $0.30 \pm 0.004$ | $2.48 \pm 0.032$ |
| PGC14520 | NGC1521 | $-5.0 \pm 0.5$ | 0.040 | $0.81 \pm 0.01$ | 1.44 | 1.31 | $17.15 \pm 0.033$ | $18.66 \pm 0.110$ | $7.41 \pm 0.13$ | $0.26 \pm 0.005$ | $2.38 \pm 0.019$ |
| PGC14757 | NGC1549 | $-5.0 \pm 0.6$ | 0.013 | $0.80 \pm 0.01$ | 1.69 | 1.90 | $14.83 \pm 0.002$ | $16.55 \pm 0.013$ | $6.84 \pm 0.08$ | $0.28 \pm 0.016$ | $2.31 \pm 0.008$ |
| PGC22322 | NGC2475 | $-5.0 \pm 0.9$ | 0.039 | $0.79 \pm 0.03$ | 0.90 | 0.95 | $18.69 \pm 0.061$ | $19.63 \pm 0.154$ | $6.74 \pm 0.25$ | – | – |
| PGC24909 | NGC2675 | $-5.0 \pm 0.8$ | 0.020 | $0.90 \pm 0.03$ | 1.18 | 1.24 | $18.84 \pm 0.016$ | $20.39 \pm 0.057$ | $7.24 \pm 0.21$ | – | $2.41 \pm 0.047$ |
| PGC25144 | NGC2693 | $-5.0 \pm 0.5$ | 0.021 | $0.82 \pm 0.01$ | 1.42 | 1.35 | $17.15 \pm 0.009$ | $18.04 \pm 0.023$ | $6.27 \pm 0.15$ | $0.32 \pm 0.004$ | $2.54 \pm 0.013$ |
| PGC26514 | NGC2810 | $-5.0 \pm 0.8$ | 0.052 | $0.84 \pm 0.02$ | 1.23 | 1.03 | $18.14 \pm 0.061$ | $18.80 \pm 0.116$ | $6.77 \pm 0.19$ | $0.30 \pm 0.006$ | $2.35 \pm 0.014$ |
| PGC29822 | NGC3158 | $-5.0 \pm 0.5$ | 0.013 | $0.88 \pm 0.01$ | 1.30 | 1.14 | $17.79 \pm 0.048$ | $18.26 \pm 0.080$ | $6.41 \pm 0.15$ | $0.32 \pm 0.004$ | $2.54 \pm 0.014$ |
| PGC29825 | NGC3159 | $-5.0 \pm 0.8$ | 0.013 | $0.85 \pm 0.02$ | 1.32 | 0.93 | $19.18 \pm 0.099$ | $20.01 \pm 0.172$ | $6.46 \pm 0.22$ | – | – |
| PGC30099 | NGC3193 | $-5.0 \pm 0.3$ | 0.026 | $0.80 \pm 0.01$ | 1.48 | 1.52 | $16.14 \pm 0.005$ | $17.63 \pm 0.018$ | $6.87 \pm 0.04$ | $0.26 \pm 0.010$ | $2.29 \pm 0.013$ |
| PGC34778 | NGC3640 | $-5.0 \pm 0.3$ | 0.044 | $0.76 \pm 0.01$ | 1.60 | 1.61 | $15.74 \pm 0.004$ | $17.78 \pm 0.022$ | $7.52 \pm 0.13$ | $0.24 \pm 0.005$ | $2.26 \pm 0.010$ |
| PGC37061 | NGC3923 | $-5.0 \pm 0.3$ | 0.083 | $0.80 \pm 0.02$ | 1.77 | 1.88 | $14.68 \pm 0.002$ | $16.10 \pm 0.017$ | $6.60 \pm 0.40$ | $0.30 \pm 0.005$ | $2.40 \pm 0.010$ |
| PGC39764 | NGC4278 | $-5.0 \pm 0.3$ | 0.029 | $0.78 \pm 0.01$ | 1.61 | 1.67 | $15.08 \pm 0.003$ | $16.01 \pm 0.009$ | $5.98 \pm 0.13$ | $0.27 \pm 0.001$ | $2.38 \pm 0.010$ |
| PGC39800 | NGC4283 | $-5.0 \pm 0.4$ | 0.025 | $0.79 \pm 0.01$ | 1.18 | 1.47 | $17.11 \pm 0.008$ | $18.52 \pm 0.027$ | $6.57 \pm 0.13$ | $0.25 \pm 0.004$ | $2.05 \pm 0.016$ |
| PGC40455 | NGC4374 | $-5.0 \pm 0.3$ | 0.040 | $0.82 \pm 0.01$ | 1.81 | 1.88 | $14.35 \pm 0.002$ | $15.60 \pm 0.006$ | $6.65 \pm 0.05$ | $0.29 \pm 0.001$ | $2.45 \pm 0.004$ |
| PGC40562 | NGC4387 | $-5.0 \pm 0.6$ | 0.033 | $0.74 \pm 0.01$ | 1.25 | 1.23 | $17.44 \pm 0.007$ | $19.09 \pm 0.027$ | $7.11 \pm 0.06$ | $0.23 \pm 0.004$ | $2.02 \pm 0.011$ |
| PGC40653 | NGC4406 | $-5.0 \pm 0.3$ | 0.030 | $0.78 \pm 0.01$ | 1.95 | 2.13 | $14.09 \pm 0.002$ | $15.43 \pm 0.005$ | $6.67 \pm 0.05$ | $0.29 \pm 0.003$ | $2.37 \pm 0.005$ |
| PGC41095 | NGC4458 | $-5.0 \pm 0.4$ | 0.024 | $0.72 \pm 0.01$ | 1.24 | 1.39 | $17.22 \pm 0.006$ | $19.38 \pm 0.028$ | $7.43 \pm 0.05$ | $0.21 \pm 0.003$ | $2.01 \pm 0.011$ |

Table 1—Continued

| PGC (1) | AName (2) | T (3) | $E(B-V)$ (4) | $(B-V)_{Tc}$ (5) | $logD_{25}$ (6) | $logD_{NUV}$ (7) | NUV (8) | FUV (9) | FUV−$V_T$ (10) | $Mg_2$ (11) | $log\sigma_V$ (12) |
|---|---|---|---|---|---|---|---|---|---|---|---|
| PGC41228 | NGC4473 | −5.0 ± 0.3 | 0.028 | 0.81 ± 0.01 | 1.65 | 1.72 | 15.43 ± 0.003 | 16.93 ± 0.009 | 6.86 ± 0.04 | 0.30 ± 0.004 | 2.25 ± 0.007 |
| PGC41963 | NGC4551 | −5.0 ± 0.6 | 0.039 | 0.79 ± 0.01 | 1.26 | 1.30 | 17.27 ± 0.006 | 18.97 ± 0.023 | 7.11 ± 0.06 | 0.25 ± 0.003 | 2.03 ± 0.009 |
| PGC41968 | NGC4552 | −5.0 ± 0.4 | 0.041 | 0.82 ± 0.01 | 1.71 | 1.78 | 14.83 ± 0.002 | 15.73 ± 0.005 | 6.15 ± 0.05 | 0.30 ± 0.005 | 2.40 ± 0.005 |
| PGC51275 | NGC5576 | −5.0 ± 0.3 | 0.031 | 0.74 ± 0.01 | 1.55 | 1.68 | 15.60 ± 0.004 | 17.55 ± 0.018 | 6.73 ± 0.13 | 0.24 ± 0.005 | 2.26 ± 0.016 |
| PGC51787 | NGC5638 | −5.0 ± 0.3 | 0.033 | 0.79 ± 0.01 | 1.43 | 1.62 | 16.27 ± 0.005 | 17.92 ± 0.021 | 6.86 ± 0.14 | 0.30 ± 0.004 | 2.22 ± 0.009 |
| PGC53643 | NGC5813 | −5.0 ± 0.3 | 0.057 | 0.81 ± 0.01 | 1.62 | 1.61 | 15.99 ± 0.010 | 17.10 ± 0.023 | 6.87 ± 0.13 | 0.29 ± 0.004 | 2.38 ± 0.008 |
| PGC61009 | NGC6482 | −5.0 ± 0.6 | 0.099 | 0.68 ± 0.01 | 1.30 | 1.14 | 16.29 ± 0.006 | 18.61 ± 0.046 | 7.51 ± 0.14 | 0.32 ± 0.003 | 2.49 ± 0.015 |
| PGC67882 | NGC7168 | −5.0 ± 0.4 | 0.023 | 0.79 ± 0.01 | 1.31 | 1.25 | 17.41 ± 0.043 | 18.94 ± 0.115 | 7.12 ± 0.17 | − | 2.26 ± 0.020 |
| PGC69256 | NGC7317 | −5.0 ± 0.5 | 0.080 | 0.77 ± 0.03 | 1.05 | 0.88 | 19.34 ± 0.028 | 20.05 ± 0.105 | 6.74 ± 0.16 | − | − |
| PGC70090 | IC1459 | −5.0 ± 0.3 | 0.016 | 0.85 ± 0.01 | 1.72 | 1.48 | 15.58 ± 0.017 | 16.49 ± 0.038 | 6.59 ± 0.16 | 0.32 ± 0.003 | 2.49 ± 0.011 |
| PGC49248 | NGC5329 | −4.7 ± 0.6 | 0.029 | 0.79 ± 0.03 | 1.12 | 1.35 | 17.91 ± 0.008 | 19.57 ± 0.039 | 7.28 ± 0.16 | − | − |
| PGC03342 | UGC579 | −4.5 ± 0.6 | 0.039 | 0.89 ± 0.01 | 1.16 | 1.05 | 19.08 ± 0.127 | 19.40 ± 0.150 | 6.22 ± 0.21 | − | 2.49 ± 0.042 |
| PGC05231 | IC1696 | −4.3 ± 0.4 | 0.042 | 0.86 ± 0.02 | 0.94 | 1.04 | 18.98 ± 0.019 | 20.46 ± 0.065 | 7.03 ± 0.16 | 0.29 ± 0.004 | 2.23 ± 0.012 |
| PGC03455 | NGC315 | −4.0 ± 0.5 | 0.065 | 0.86 ± 0.02 | 1.51 | 1.30 | 16.84 ± 0.021 | 17.58 ± 0.047 | 6.66 ± 0.21 | 0.30 ± 0.004 | 2.43 ± 0.044 |
| PGC04224 | NGC410 | −4.0 ± 0.6 | 0.058 | 0.86 ± 0.01 | 1.38 | 1.41 | 17.08 ± 0.010 | 17.93 ± 0.029 | 6.67 ± 0.13 | 0.34 ± 0.003 | 2.48 ± 0.010 |
| PGC04363 | NGC426 | −4.0 ± 0.5 | 0.032 | 0.80 ± 0.04 | 1.14 | 0.81 | 17.95 ± 0.051 | 18.67 ± 0.109 | 5.96 ± 0.42 | − | 2.44 ± 0.062 |
| PGC05766 | NGC596 | −4.0 ± 0.5 | 0.058 | 0.72 ± 0.01 | 1.51 | 1.54 | 15.93 ± 0.005 | 17.77 ± 0.021 | 7.06 ± 0.13 | 0.23 ± 0.003 | 2.18 ± 0.012 |
| PGC58265 | NGC6166 | −4.0 ± 0.4 | 0.012 | 0.87 ± 0.04 | 1.29 | 1.51 | 17.41 ± 0.009 | 18.12 ± 0.023 | 6.42 ± 0.12 | 0.32 ± 0.004 | 2.50 ± 0.008 |
| PGC65379 | NGC6964 | −4.0 ± 0.5 | 0.094 | 0.80 ± 0.01 | 1.23 | 1.38 | 17.52 ± 0.009 | 18.64 ± 0.054 | 5.99 ± 0.21 | − | 2.34 ± 0.027 |
| PGC65436 | NGC6958 | −3.8 ± 0.3 | 0.045 | 0.75 ± 0.01 | 1.33 | 1.45 | 16.42 ± 0.010 | 18.40 ± 0.042 | 7.16 ± 0.14 | 0.23 ± 0.006 | 2.28 ± 0.020 |
| PGC05326 | NGC548 | −3.7 ± 0.7 | 0.037 | 0.79 ± 0.02 | 0.95 | 1.20 | 18.95 ± 0.018 | 20.57 ± 0.069 | 7.03 ± 0.17 | 0.25 ± 0.005 | 2.17 ± 0.018 |
| PGC12922 | Haro 20 | −3.7 ± 0.7 | 0.039 | 0.36 ± 0.04 | 1.02 | 0.91 | 16.29 ± 0.021 | 16.67 ± 0.044 | 2.46 ± 0.21 | − | − |
| PGC13360 | NGC1389 | −3.3 ± 0.4 | 0.011 | 0.79 ± 0.01 | 1.36 | 1.38 | 16.78 ± 0.006 | 18.63 ± 0.025 | 7.21 ± 0.13 | − | 2.14 ± 0.026 |
| PGC02253 | NGC179 | −3.0 ± 0.6 | 0.020 | 0.80 ± 0.01 | 0.97 | 0.99 | 18.38 ± 0.054 | 19.69 ± 0.162 | 6.55 ± 0.21 | 0.28 ± 0.008 | 2.38 ± 0.033 |
| PGC04042 | NGC392 | −3.0 ± 1.2 | 0.058 | 0.79 ± 0.01 | 1.08 | 1.17 | 17.85 ± 0.014 | 18.85 ± 0.044 | 6.36 ± 0.14 | 0.29 ± 0.004 | 2.41 ± 0.012 |
| PGC05305 | NGC541 | −3.0 ± 0.4 | 0.045 | 0.79 ± 0.03 | 1.25 | 1.34 | 17.64 ± 0.010 | 18.87 ± 0.032 | 6.97 ± 0.16 | 0.31 ± 0.005 | 2.32 ± 0.010 |
| PGC05311 | NGC543 | −3.0 ± 0.8 | 0.040 | 0.88 ± 0.04 | 0.77 | 1.02 | 18.94 ± 0.018 | 19.80 ± 0.048 | 6.90 ± 0.21 | 0.32 ± 0.010 | 2.38 ± 0.014 |
| PGC05468 | NGC568 | −3.0 ± 0.5 | 0.019 | 0.85 ± 0.03 | 1.35 | 1.35 | 17.96 ± 0.011 | 19.09 ± 0.032 | 6.60 ± 0.18 | − | 2.32 ± 0.025 |
| PGC05565 | - | −3.0 ± 0.8 | 0.018 | 0.89 ± 0.03 | 1.16 | 1.08 | 19.12 ± 0.073 | 20.20 ± 0.190 | 6.93 ± 0.28 | − | − |
| PGC06957 | NGC703 | −3.0 ± 1.2 | 0.091 | 0.80 ± 0.03 | 1.08 | 0.83 | 18.73 ± 0.104 | 19.21 ± 0.155 | 6.28 ± 0.22 | 0.31 ± 0.008 | 2.44 ± 0.029 |
| PGC07763 | NGC794 | −3.0 ± 1.1 | 0.063 | 0.90 ± 0.03 | 1.11 | 0.96 | 18.43 ± 0.086 | 19.10 ± 0.140 | 6.61 ± 0.21 | − | 2.44 ± 0.033 |
| PGC10123 | NGC1023 | −3.0 ± 0.3 | 0.061 | 0.82 ± 0.01 | 1.94 | 1.77 | 14.87 ± 0.003 | 16.40 ± 0.016 | 7.28 ± 0.06 | 0.27 ± 0.002 | 2.31 ± 0.009 |
| PGC10302 | NGC1060 | −3.0 ± 1.0 | 0.209 | 0.86 ± 0.01 | 1.36 | 1.24 | 16.78 ± 0.016 | 17.63 ± 0.060 | 6.51 ± 0.15 | 0.31 ± 0.006 | 2.48 ± 0.024 |
| PGC11774 | NGC1222 | −3.0 ± 1.2 | 0.060 | 0.42 ± 0.01 | 1.04 | 1.11 | 14.98 ± 0.013 | 15.58 ± 0.029 | 3.31 ± 0.13 | − | − |

Table 1—Continued

| PGC (1) | AName (2) | T (3) | E(B−V) (4) | (B−V)$_{Tc}$ (5) | logD$_{25}$ (6) | logD$_{NUV}$ (7) | NUV (8) | FUV (9) | FUV−V$_T$ (10) | Mg$_2$ (11) | logσ$_V$ (12) |
|---|---|---|---|---|---|---|---|---|---|---|---|
| PGC13344 | NGC1387 | −3.0 ± 0.3 | 0.013 | 0.86 ± 0.01 | 1.45 | 1.49 | 15.57 ± 0.003 | 16.54 ± 0.008 | 5.93 ± 0.10 | − | − |
| PGC13470 | NGC1400 | −3.0 ± 0.3 | 0.065 | 0.78 ± 0.01 | 1.36 | 1.40 | 15.86 ± 0.005 | 16.72 ± 0.015 | 6.00 ± 0.13 | 0.30 ± 0.004 | 2.41 ± 0.010 |
| PGC21896 | UGC4014 | −3.0 ± 1.3 | 0.025 | 0.66 ± 0.03 | 0.92 | 0.88 | 17.60 ± 0.036 | 18.42 ± 0.090 | 4.94 ± 0.22 | − | − |
| PGC24707 | NGC2656 | −3.0 ± 0.8 | 0.026 | 1.00 ± 0.04 | 1.12 | 1.05 | 18.95 ± 0.017 | 19.97 ± 0.048 | 6.34 ± 0.21 | − | − |
| PGC29846 | NGC3163 | −3.0 ± 0.7 | 0.013 | 0.89 ± 0.02 | 1.17 | 0.98 | 18.96 ± 0.087 | 19.92 ± 0.166 | 6.69 ± 0.22 | − | 2.33 ± 0.052 |
| PGC41538 | NGC4503 | −3.0 ± 0.4 | 0.050 | 0.81 ± 0.01 | 1.55 | 1.57 | 16.42 ± 0.007 | 18.05 ± 0.016 | 7.18 ± 0.08 | − | 2.05 ± 0.090 |
| PGC41634 | IC3481 | −3.0 ± 0.6 | 0.056 | 0.86 ± 0.02 | 0.92 | 1.22 | 18.45 ± 0.024 | 19.28 ± 0.029 | 6.15 ± 0.15 | − | − |
| PGC41939 | NGC4546 | −3.0 ± 0.3 | 0.034 | 0.83 ± 0.02 | 1.52 | 1.62 | 15.61 ± 0.005 | 17.39 ± 0.018 | 7.22 ± 0.13 | − | 2.38 ± 0.048 |
| PGC43981 | NGC4798 | −3.0 ± 1.2 | 0.012 | 0.90 ± 0.02 | 1.08 | 1.22 | 18.43 ± 0.014 | 19.82 ± 0.050 | 6.73 ± 0.21 | − | − |
| PGC51270 | NGC5574 | −3.0 ± 0.9 | 0.031 | 0.69 ± 0.01 | 1.21 | 1.36 | 16.98 ± 0.007 | 19.35 ± 0.041 | 7.10 ± 0.14 | − | 1.88 ± 0.020 |
| PGC66113 | - | −3.0 ± 0.5 | 0.032 | 0.64 ± 0.01 | 1.16 | 1.10 | 17.48 ± 0.037 | 19.09 ± 0.128 | 6.34 ± 0.18 | − | − |
| PGC67303 | NGC7117 | −3.0 ± 0.6 | 0.028 | 0.82 ± 0.02 | 1.15 | 0.96 | 18.69 ± 0.063 | 19.74 ± 0.167 | 7.08 ± 0.21 | 0.29 ± 0.010 | 2.41 ± 0.055 |
| PGC71518 | NGC7675 | −3.0 ± 0.6 | 0.058 | 0.92 ± 0.03 | 0.82 | 0.99 | 19.16 ± 0.022 | 20.75 ± 0.078 | 6.11 ± 0.17 | − | 2.26 ± 0.036 |
| PGC13354 | NGC1382 | −2.7 ± 0.6 | 0.017 | 0.81 ± 0.02 | 1.18 | 1.55 | 17.38 ± 0.006 | 18.92 ± 0.027 | 7.04 ± 0.13 | 0.17 ± 0.035 | 1.98 ± 0.065 |
| PGC67318 | NGC7118 | −2.7 ± 0.6 | 0.030 | 0.83 ± 0.01 | 1.17 | 1.06 | 18.36 ± 0.066 | 19.07 ± 0.152 | 6.65 ± 0.20 | 0.26 ± 0.010 | 2.28 ± 0.055 |
| PGC02359 | NGC193 | −2.5 ± 0.6 | 0.023 | 0.90 ± 0.02 | 1.16 | 1.15 | 18.03 ± 0.045 | 19.11 ± 0.122 | 7.01 ± 0.19 | − | − |
| PGC13445 | NGC1403 | −2.5 ± 0.6 | 0.025 | 0.77 ± 0.01 | 1.13 | 1.21 | 17.86 ± 0.057 | 19.23 ± 0.129 | 6.61 ± 0.18 | 0.27 ± 0.007 | 2.22 ± 0.014 |
| PGC28974 | NGC3073 | −2.5 ± 0.6 | 0.010 | 0.54 ± 0.02 | 1.11 | 1.22 | 16.70 ± 0.006 | 17.50 ± 0.015 | 4.18 ± 0.14 | − | 1.54 ± 0.046 |
| PGC40484 | NGC4379 | −2.5 ± 0.5 | 0.023 | 0.77 ± 0.01 | 1.28 | 1.29 | 16.83 ± 0.007 | 18.34 ± 0.025 | 6.73 ± 0.07 | − | 2.04 ± 0.016 |
| PGC60025 | NGC6359 | −2.5 ± 0.6 | 0.021 | 0.78 ± 0.02 | 1.09 | 1.22 | 17.74 ± 0.008 | 18.95 ± 0.035 | 6.42 ± 0.14 | 0.28 ± 0.009 | 2.27 ± 0.046 |
| PGC68826 | IC1445 | −2.5 ± 0.5 | 0.041 | 0.68 ± 0.03 | 1.21 | 1.06 | 18.21 ± 0.057 | 20.68 ± 0.271 | 8.19 ± 0.34 | − | − |
| PGC69964 | NGC7404 | −2.5 ± 0.9 | 0.012 | 0.73 ± 0.02 | 1.17 | 1.25 | 17.56 ± 0.042 | 19.85 ± 0.161 | 8.05 ± 0.21 | − | − |
| PGC03367 | NGC307 | −2.3 ± 0.7 | 0.043 | 0.81 ± 0.02 | 1.20 | 0.90 | 18.33 ± 0.057 | 19.94 ± 0.192 | 7.34 ± 0.23 | − | 2.49 ± 0.020 |
| PGC14375 | NGC1510 | −2.3 ± 0.7 | 0.011 | 0.32 ± 0.01 | 1.12 | 1.61 | 14.63 ± 0.002 | 14.89 ± 0.004 | 1.95 ± 0.11 | − | − |
| PGC41781 | NGC4528 | −2.3 ± 0.7 | 0.045 | 0.75 ± 0.01 | 1.22 | 1.31 | 17.01 ± 0.008 | 18.92 ± 0.024 | 7.04 ± 0.09 | − | 2.07 ± 0.026 |
| PGC43895 | NGC4789 | −2.3 ± 0.7 | 0.008 | 0.89 ± 0.03 | 1.28 | 1.33 | 18.07 ± 0.012 | 19.22 ± 0.037 | 7.19 ± 0.16 | 0.29 ± 0.004 | 2.43 ± 0.008 |
| PGC02603 | IC48 | −2.2 ± 0.7 | 0.038 | 0.69 ± 0.01 | 0.99 | 1.10 | 16.48 ± 0.026 | 16.83 ± 0.048 | 3.86 ± 0.14 | − | 2.17 ± 0.041 |
| PGC72173 | NGC7736 | −2.2 ± 0.5 | 0.028 | 0.84 ± 0.02 | 1.22 | 1.11 | 18.47 ± 0.081 | 19.82 ± 0.175 | 7.05 ± 0.25 | − | − |
| PGC00243 | NGC7808 | −2.0 ± 1.2 | 0.043 | 0.69 ± 0.03 | 1.10 | 1.15 | 16.86 ± 0.012 | 17.46 ± 0.028 | 5.01 ± 0.14 | − | − |
| PGC02397 | NGC204 | −2.0 ± 0.8 | 0.023 | 0.77 ± 0.02 | 1.08 | 1.04 | 18.22 ± 0.049 | 19.66 ± 0.155 | 6.88 ± 0.25 | − | − |
| PGC02959 | NGC273 | −2.0 ± 0.4 | 0.057 | 0.76 ± 0.02 | 1.34 | 1.18 | 18.02 ± 0.069 | 18.84 ± 0.125 | 6.13 ± 0.19 | − | − |
| PGC03434 | NGC311 | −2.0 ± 0.8 | 0.066 | 0.81 ± 0.02 | 1.18 | 1.09 | 18.38 ± 0.044 | 19.21 ± 0.098 | 6.45 ± 0.16 | − | 2.45 ± 0.024 |
| PGC04793 | NGC471 | −2.0 ± 0.9 | 0.051 | 0.68 ± 0.01 | 1.02 | 0.93 | 17.21 ± 0.009 | 18.34 ± 0.026 | 5.21 ± 0.13 | − | 2.04 ± 0.204 |
| PGC04801 | NGC474 | −2.0 ± 0.3 | 0.034 | 0.71 ± 0.02 | 1.85 | 1.65 | 16.23 ± 0.005 | 17.99 ± 0.021 | 6.63 ± 0.13 | − | 2.21 ± 0.013 |

Table 1—Continued

| PGC (1) | AName (2) | T (3) | $E(B-V)$ (4) | $(B-V)_{Tc}$ (5) | $logD_{25}$ (6) | $logD_{NUV}$ (7) | NUV (8) | FUV (9) | FUV–$V_T$ (10) | Mg$_2$ (11) | $log\sigma_V$ (12) |
|---|---|---|---|---|---|---|---|---|---|---|---|
| PGC06570 | NGC670 | $-2.0 \pm 0.6$ | 0.072 | $0.68 \pm 0.01$ | 1.31 | 1.19 | $15.44 \pm 0.016$ | $15.79 \pm 0.032$ | $3.34 \pm 0.13$ | $0.19 \pm 0.010$ | $2.12 \pm 0.050$ |
| PGC06782 | NGC687 | $-2.0 \pm 0.8$ | 0.062 | $0.86 \pm 0.01$ | 1.15 | 1.09 | $18.06 \pm 0.076$ | $19.49 \pm 0.157$ | $7.47 \pm 0.20$ | $0.30 \pm 0.004$ | $2.38 \pm 0.011$ |
| PGC08258 | NGC842 | $-2.0 \pm 0.6$ | 0.024 | $0.81 \pm 0.01$ | 1.09 | 1.34 | $17.82 \pm 0.010$ | $19.40 \pm 0.037$ | $6.86 \pm 0.16$ | – | – |
| PGC12651 | NGC1316 | $-2.0 \pm 0.3$ | 0.021 | $0.75 \pm 0.01$ | 2.08 | 2.09 | $13.58 \pm 0.001$ | $15.32 \pm 0.006$ | $6.90 \pm 0.08$ | $0.23 \pm 0.015$ | $2.36 \pm 0.008$ |
| PGC13318 | NGC1380 | $-2.0 \pm 0.6$ | 0.017 | $0.80 \pm 0.01$ | 1.68 | 1.73 | $15.29 \pm 0.002$ | $16.78 \pm 0.009$ | $6.94 \pm 0.10$ | $0.25 \pm 0.008$ | $2.35 \pm 0.011$ |
| PGC13335 | NGC1380A | $-2.0 \pm 0.8$ | 0.015 | $0.77 \pm 0.01$ | 1.38 | 1.47 | $17.26 \pm 0.006$ | $19.21 \pm 0.029$ | $6.89 \pm 0.13$ | – | $1.83 \pm 0.030$ |
| PGC13377 | NGC1383 | $-2.0 \pm 0.6$ | 0.072 | $0.79 \pm 0.01$ | 1.28 | 1.21 | $17.80 \pm 0.013$ | $19.34 \pm 0.063$ | $7.14 \pm 0.15$ | – | $2.25 \pm 0.030$ |
| PGC13444 | NGC1394 | $-2.0 \pm 0.8$ | 0.073 | $0.82 \pm 0.02$ | 1.13 | 1.20 | $17.64 \pm 0.012$ | $18.81 \pm 0.042$ | $6.27 \pm 0.14$ | – | – |
| PGC13467 | NGC1402 | $-2.0 \pm 0.8$ | 0.064 | $0.61 \pm 0.02$ | 0.91 | 0.91 | $17.12 \pm 0.009$ | $18.20 \pm 0.028$ | $4.88 \pm 0.13$ | – | – |
| PGC14765 | NGC1553 | $-2.0 \pm 0.3$ | 0.013 | $0.75 \pm 0.01$ | 1.65 | 1.84 | $14.53 \pm 0.002$ | $16.18 \pm 0.010$ | $6.86 \pm 0.08$ | $0.26 \pm 0.005$ | $2.25 \pm 0.010$ |
| PGC25950 | NGC2784 | $-2.0 \pm 0.3$ | 0.214 | $0.81 \pm 0.01$ | 1.74 | 1.50 | $15.08 \pm 0.006$ | $16.20 \pm 0.038$ | $6.75 \pm 0.14$ | $0.29 \pm 0.004$ | – |
| PGC27765 | NGC2950 | $-2.0 \pm 0.3$ | 0.017 | $0.76 \pm 0.02$ | 1.43 | 1.25 | $16.38 \pm 0.022$ | $18.20 \pm 0.079$ | $7.35 \pm 0.15$ | – | – |
| PGC38742 | NGC4150 | $-2.0 \pm 0.4$ | 0.018 | $0.66 \pm 0.01$ | 1.37 | 1.30 | $16.17 \pm 0.007$ | $17.89 \pm 0.025$ | $6.35 \pm 0.13$ | $0.09 \pm 0.005$ | $1.93 \pm 0.041$ |
| PGC40898 | NGC4435 | $-2.0 \pm 0.3$ | 0.030 | $0.79 \pm 0.01$ | 1.44 | 1.79 | $15.55 \pm 0.003$ | $17.12 \pm 0.010$ | $6.45 \pm 0.05$ | $0.19 \pm 0.003$ | $2.19 \pm 0.016$ |
| PGC41260 | NGC4477 | $-2.0 \pm 0.4$ | 0.032 | $0.81 \pm 0.01$ | 1.58 | 1.63 | $15.74 \pm 0.003$ | $16.98 \pm 0.010$ | $6.70 \pm 0.05$ | – | $2.27 \pm 0.034$ |
| PGC41302 | NGC4479 | $-2.0 \pm 0.6$ | 0.029 | $0.81 \pm 0.01$ | 1.19 | 1.35 | $17.44 \pm 0.007$ | $18.84 \pm 0.022$ | $6.53 \pm 0.06$ | $0.18 \pm 0.011$ | $1.92 \pm 0.025$ |
| PGC42149 | NGC4578 | $-2.0 \pm 0.5$ | 0.021 | $0.72 \pm 0.02$ | 1.52 | 1.23 | $17.27 \pm 0.046$ | $18.63 \pm 0.144$ | $7.22 \pm 0.16$ | $0.28 \pm 0.013$ | $2.08 \pm 0.033$ |
| PGC53201 | NGC5770 | $-2.0 \pm 0.8$ | 0.038 | $0.73 \pm 0.01$ | 1.23 | 1.31 | $16.99 \pm 0.005$ | $19.20 \pm 0.041$ | $7.10 \pm 0.14$ | – | $2.07 \pm 0.127$ |
| PGC65376 | NGC6963 | $-2.0 \pm 0.9$ | 0.095 | $0.79 \pm 0.04$ | 0.82 | 1.00 | $18.58 \pm 0.015$ | $19.62 \pm 0.064$ | $5.96 \pm 0.21$ | – | – |
| PGC68606 | NGC7249 | $-2.0 \pm 0.7$ | 0.021 | $0.86 \pm 0.04$ | 1.05 | 0.81 | $19.91 \pm 0.105$ | $20.93 \pm 0.263$ | $7.65 \pm 0.31$ | – | – |
| PGC68612 | NGC7252 | $-2.0 \pm 0.5$ | 0.030 | $0.51 \pm 0.01$ | 1.29 | 1.46 | $15.22 \pm 0.005$ | $16.48 \pm 0.017$ | $5.21 \pm 0.13$ | – | $2.20 \pm 0.053$ |
| PGC71478 | NGC7671 | $-2.0 \pm 0.6$ | 0.073 | $0.81 \pm 0.01$ | 1.14 | 1.00 | $18.21 \pm 0.081$ | $19.52 \pm 0.178$ | $6.99 \pm 0.22$ | – | – |
| PGC71554 | NGC7679 | $-2.0 \pm 0.4$ | 0.065 | $0.36 \pm 0.01$ | 1.13 | 1.08 | $14.71 \pm 0.012$ | $15.39 \pm 0.027$ | $2.71 \pm 0.13$ | – | – |
| PGC71565 | NGC7683 | $-2.0 \pm 0.8$ | 0.070 | $0.87 \pm 0.01$ | 1.28 | 1.15 | $18.03 \pm 0.080$ | $19.45 \pm 0.172$ | $7.17 \pm 0.26$ | – | – |
| PGC13425 | NGC1393 | $-1.9 \pm 0.4$ | 0.067 | $0.76 \pm 0.02$ | 1.29 | 1.24 | $17.38 \pm 0.010$ | $18.63 \pm 0.041$ | $6.86 \pm 0.14$ | – | $2.06 \pm 0.030$ |
| PGC13091 | NGC1352 | $-1.8 \pm 0.6$ | 0.038 | $0.78 \pm 0.02$ | 1.02 | 1.00 | $18.85 \pm 0.102$ | $20.94 \pm 0.280$ | $7.77 \pm 0.31$ | – | – |
| PGC39922 | NGC4292 | $-1.7 \pm 0.5$ | 0.021 | $0.74 \pm 0.02$ | 1.22 | 1.37 | $16.99 \pm 0.007$ | $18.48 \pm 0.025$ | $6.38 \pm 0.15$ | – | $1.76 \pm 0.022$ |
| PGC13321 | NGC1381 | $-1.6 \pm 0.5$ | 0.013 | $0.81 \pm 0.01$ | 1.43 | 1.46 | $16.75 \pm 0.005$ | $18.31 \pm 0.018$ | $6.90 \pm 0.10$ | $0.25 \pm 0.020$ | $2.18 \pm 0.011$ |

Note. — Col.(1&2): PGC number and alternate name. Col.(3): morphological type from RC3. Col.(4): Color excess in magnitude from the Schlegel et al. (1998) reddening maps, used for the Galactic extinction corrections. Col.(5): Total (asymptotic) color index from RC3, converted in the AB system and corrected for Galactic extinction. Col.(6): Decimal logarithm of the apparent major isophotal diameter from RC3. Unit of $D_{25}$ is $0\rlap{.}'1$. Col.(7): Decimal logarithm of the major diameter of the elliptical aperture determined by the GALEX pipeline in the near-UV image. Unit of $D_{NUV}$ is $0\rlap{.}'1$. Col.(8): near-UV magnitude (in the AB system) given by the GALEX

pipeline in the elliptical aperture $D_{NUV}$ (col.7) and corrected for Galactic extinction. Col.(9): far-UV magnitude (in the AB system) in the aperture $D_{NUV}$ used for the near-UV magnitude integration, and corrected for Galactic extinction. Col.(10): FUV$-V_T$ color calculated with FUV$-V_T$ = FUV+$(B-V)_{Tc} - B_{Tc}$, where $B_{Tc}$ is the $B_T$ from RC3 converted in the AB system and corrected for Galactic extinction. Col.(11): Mg$_2$ index from the compilation of Golev & Prugniel (1998). Col.(12): Decimal logarithm of the central velocity dispersion $\sigma_V$ in kms$^{-1}$ from the LEDA database.

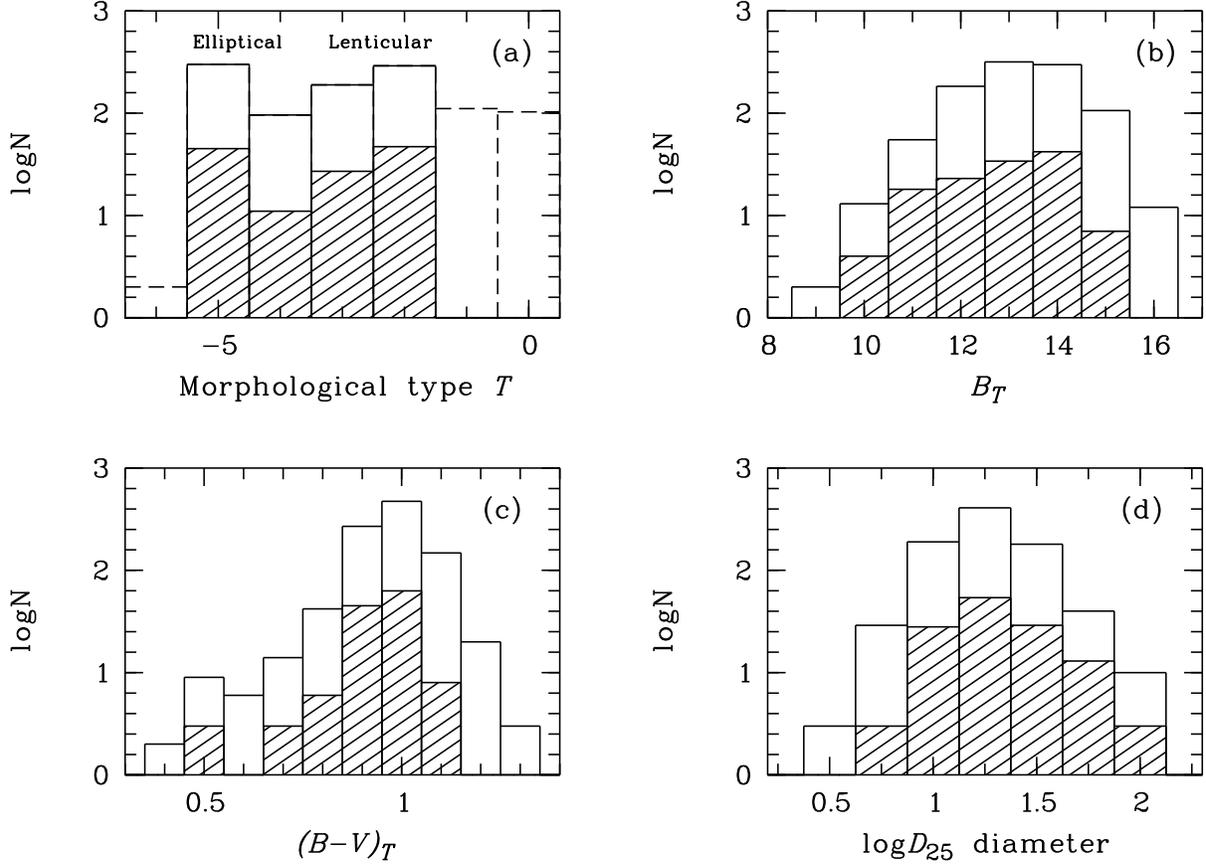

Fig. 1.— Characteristics of our sample. Numbers of galaxies (in logN) in our sample (hatched histogram) compared to the early-type galaxies ($-5.5 < T < -1.5$) in the RC3 catalog entries having $(B - V)_T$ measurements in bins of morphological type $T$ (a), $B_T$ magnitudes (b), $(B - V)_T$ colors (c), and $D_{25}$ diameter (d). The $B_T$ and $(B - V)_T$ are in the magnitude system adopted in the RC3.

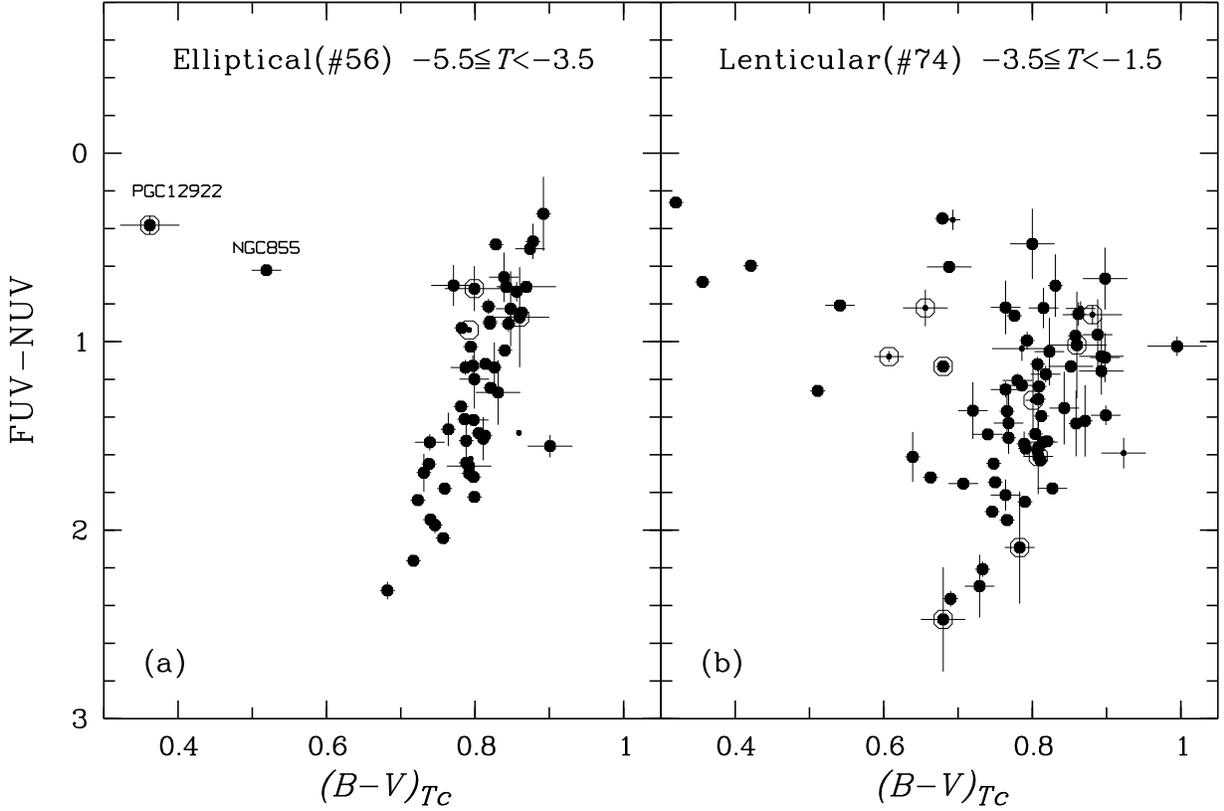

Fig. 2.— Color-color diagrams FUV–NUV vs. $(B-V)_{Tc}$ for the Elliptical and Lenticular subsamples. The Elliptical subsample (*left panel*) shows a clear anti-correlation. Two galaxies, PGC12922 and PGC8557 (NCG855) are off the correlation. NGC855 appears similar to an E6 or E7 galaxy on the POSS plate, whereas HST observation of the inner structure clearly shows features of an edge-on late-type galaxy (Phillips et al. 1996), and Walsh et al. (1990) find an extended HI distribution using the VLA. PGC12922 is the peculiar blue galaxy Haro 20. The Lenticular subsample (*right panel*) shows a more dispersed anti-correlation and a larger fraction of galaxies off the correlation. Open circles indicate the galaxies for which the near–UV or far–UV flux contribution in a central $6''$ aperture is larger than 30%. Small filled circles indicate galaxies with major isophotal diameter $\log D_{25} < 1$ (i.e. major isophotal diameter $< 1'$). We conclude that we are seeing an extension of the color-metallicity relationship for early-type galaxies into the UV (see text). The larger scatter and greater number of outliers suggests that lenticulars have a more complex star formation history.

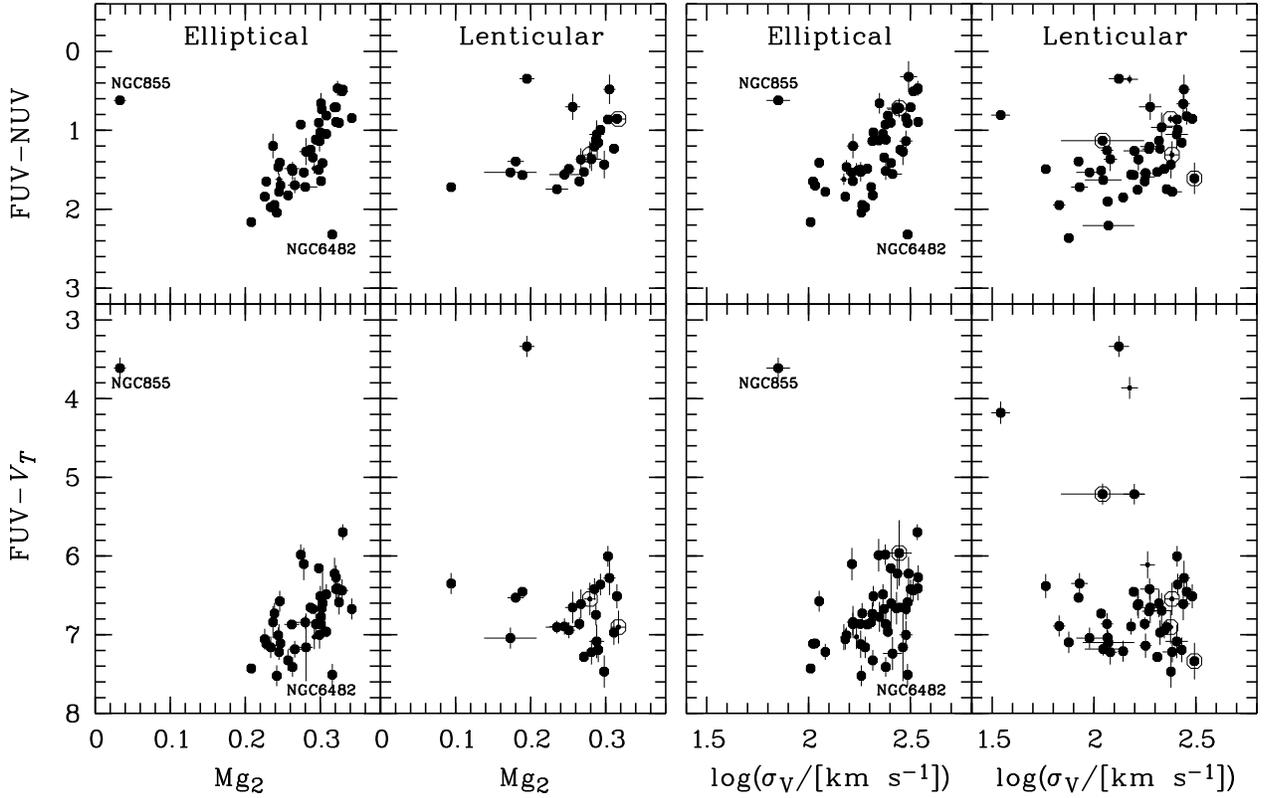

Fig. 3.— FUV−NUV and FUV−$V_T$ colors vs. Mg$_2$ index and central velocity dispersion $\sigma_V$. The Mg$_2$ and $\sigma_V$ values come from the compilation of Golev & Prugniel, and LEDA database respectively. The Elliptical subsample shows an anti-correlation (FUV−NUV color decreases when Mg$_2$ or $\sigma_V$ increases) which are reversed in sense from the well-known dependence of $B − V$ color on metal abundance or central velocity dispersion. This correlation gets worse either with the FUV−$V_T$ color, or the Lenticular subsample. Same symbols as Fig.2.

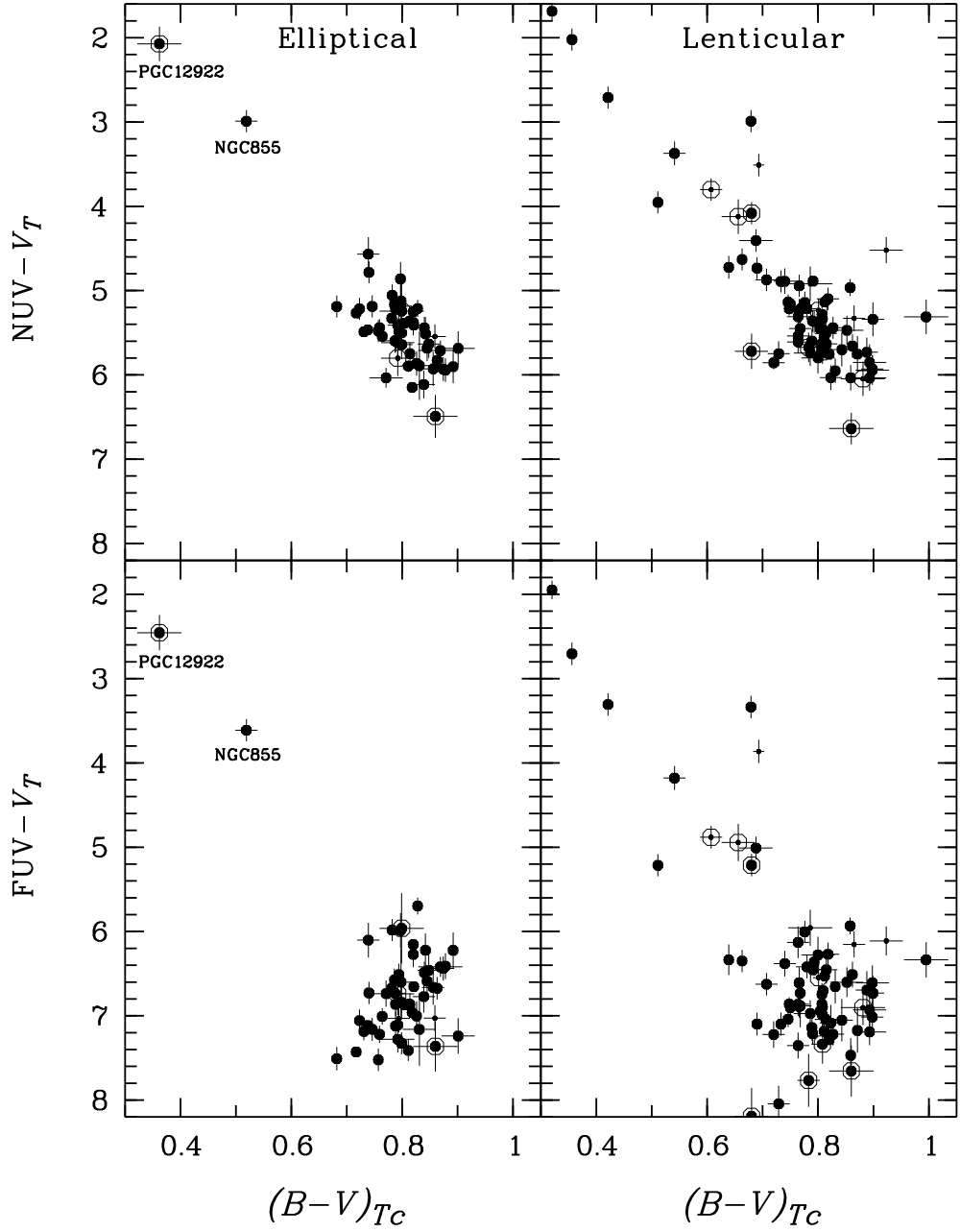

Fig. 4.— Color-color diagrams NUV–$V_T$ and FUV–$V_T$ vs. $(B-V)_{Tc}$. Same symbols as Fig.2.

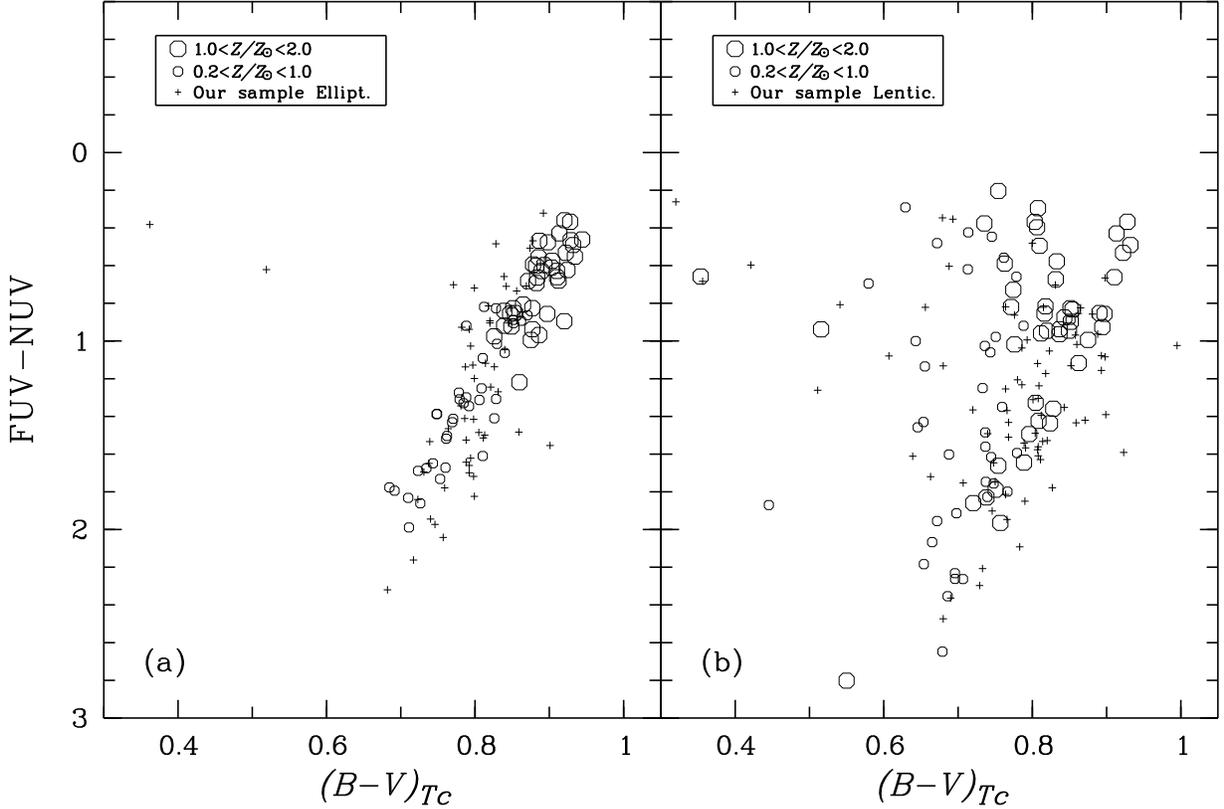

Fig. 5.— Comparison of the observed colors (*crosses*) as displayed in Fig.2, with the stochastic realizations of model SF histories as generated by Salim et al. (2005) from the Bruzual & Charlot (2003) population synthesis models (*empty circles*). *Left panel*: stochastic realizations representative of old galaxies: exponentially declining Star Formation law [SFR$(t) \propto exp(-\gamma t)$] with $\gamma > 0.9$, mean age of star formation episodes in the 8–12 Gyr interval, and low extinction (effective $V$-band absorption optical depth $\tau_V < 0.5$). *Right panel*: same selection but with mean age of star formation episodes in the 4–12 Gyr interval. The models well reproduce both the tight anti-correlation between the UV and $B - V$ colors for the Elliptical subsample (*left panel*), and the scatter for the Lenticular subsample (*right panel*).